# Accepted in PET Clinics 2025

# An overview of PBPK and PopPK Models: Applications to Radiopharmaceutical Therapies for Analysis and Personalization

Deni Hardiansyah[a*], Bisma Barron Patrianesha[a,b], Kuangyu Shi[c], Babak Saboury[d], Arman Rahmim[e], Gerhard Glatting[f]

[a]*Medical Physics and Biophysics, Physics Department, Faculty of Mathematics and Natural Sciences, Universitas Indonesia, Depok, Indonesia;*
[b]*Research Center for Safety, Metrology, and Nuclear Quality Technology, National Research and Innovation Agency, Tangerang Selatan, Indonesia;*
[c]*Department of Nuclear Medicine, Bern University Hospital, University of Bern, Bern, Switzerland;*
[d]*United Theranostics, Bethesda, Maryland, United States;*
[e]*Departments of Radiology and Physics, University of British Columbia, Vancouver, British Columbia, Canada;*
[f]*Medical Radiation Physics, Department of Nuclear Medicine, Ulm University, Ulm, Germany*

*Corresponding Author:

Dr. sc. hum. Deni Hardiansyah
Physics Department, Faculty of Mathematics and Natural Sciences (FMIPA),
Universitas Indonesia, Depok, 16424
Email (University): denihardiansyah@ui.ac.id

**Key Points:**

- Physiologically based pharmacokinetic (PBPK) models predict radiopharmaceutical distribution by integrating physiological and drug parameters as prior knowledge, enabling personalized dosing.
- Population pharmacokinetic (PopPK) models utilize clinical data to account for interindividual and intraindividual variability, optimizing dose estimates even with limited patient data.
- Hybrid PBPK-PopPK modeling is poised to considerably advance personalized radiopharmaceutical therapy.
- Robust validation and standardization are essential for clinical implementation and regulatory approval of pharmacokinetic models.

**Synopsis:** This review discusses the current applications, advantages, and limitations of PBPK and PopPK models in radiopharmaceutical therapy (RPT). PBPK models simulate radiopharmaceutical kinetics by integrating prior physiological and drug parameter information, whereas PopPK models leverage population data to enhance individual dose estimation accuracy. Future directions include developing hybrid models, incorporating artificial intelligence, and establishing regulatory guidelines to promote their clinical adoption. Ultimately, these modeling strategies aim to enable precise, personalized RPT dosing, thereby improving therapeutic outcomes and safety.

## Introduction

Accurate individualized absorbed dose estimation is increasingly regarded as an essential prerequisite for optimizing therapeutic efficacy in radiopharmaceutical therapy (RPT), while simultaneously ensuring that the radiation burden to healthy tissues remains within tolerable limits.[1-6] Absorbed dose estimation relies primarily on two key parameters: the time-integrated activity coefficient (TIAC) and the dose conversion factor.[7] By quantifying the absorbed dose delivered to organs-at-risk and tumor lesions, treatment protocols—including the determination of the optimal administered activity—can be tailored to the physiology of each patient, supporting more effective and safer therapy.

One of the primary clinical objectives of dosimetry is to individualize treatment and reduce interpatient variability in absorbed doses, thereby improving the therapeutic response and minimizing the risk of adverse effects. Without individualized adjustments, the absorbed doses to critical organs at risk can exhibit substantial inter-patient variability, often spanning up to one order of magnitude.[8] For example, kidney absorbed doses for [$^{177}$Lu]Lu-PSMA-I&T have shown a coefficient of variation (CV) of approximately 22% to 60% in Gy/GBq. Similarly, for [$^{177}$Lu]Lu-DOTATATE, marked inter-individual variability has been reported, with kidney absorbed doses ranging from 0.33 to 2.4 Gy/GBq (mean = 0.80 Gy/GBq, SD = 0.30).[9,10] This underscores the importance of accurately identifying and individually measuring key determinants of absorbed dose. Among the determinants of absorbed dose, TIAC shows substantially greater inter-individual variability than the dose conversion factor[11], making patient-specific TIAC estimation particularly critical for achieving reliable absorbed dose calculations. Accurate individualized biokinetic modeling of time-activity curves (TACs) to calculate TIACs is therefore essential to guide therapy optimization and ultimately improve patient outcomes.[12,13]

The estimation of TACs for dosimetry typically involves four key components.[13,14] The first component is data input, where quantitative measurements from imaging or blood sampling are collected. The second component, function input, defines the mathematical model or structure used to describe the temporal changes in radiopharmaceutical activity. The third component, data processing, includes curve fitting procedures and the quantification of associated uncertainties. The fourth component involves presenting the output data, such as the TIAC and fitted parameter estimates. The overall accuracy of TIAC — and, consequently, absorbed dose estimation — can be improved by incorporating prior knowledge at both the function input and data processing components.[15-17]

Several strategies are available to define the function input, each with varying degrees of complexity and predictive capability. Simple trapezoidal integration and extrapolations offer straightforward numerical estimation of TIAC. However, these methods are prone to under- or over-estimation, particularly when extrapolated to infinite time.[18] To overcome these limitations, analytical functions such as sum-of-exponential functions (SOEF) are commonly employed.[7,18-21] SOEFs provide a more structured description of biokinetic behavior and facilitate better extrapolations beyond the measured data points. Another method of function input is the use of compartmental models.[18,22] More complex yet clinically relevant are physiologically based pharmacokinetic (PBPK) models, which define compartments based on anatomical and physiological volumes connected by blood flows and clearance, as well as the drug characteristics.[23-25] By using the model structure with biological and drug prior knowledge, PBPK models enable predictive simulations of radiopharmaceutical distribution and clearance, even under conditions not observed in the clinical data.[26] This predictive capability is particularly valuable in clinical practice, where imaging schedules are often limited to reduce patient burden.

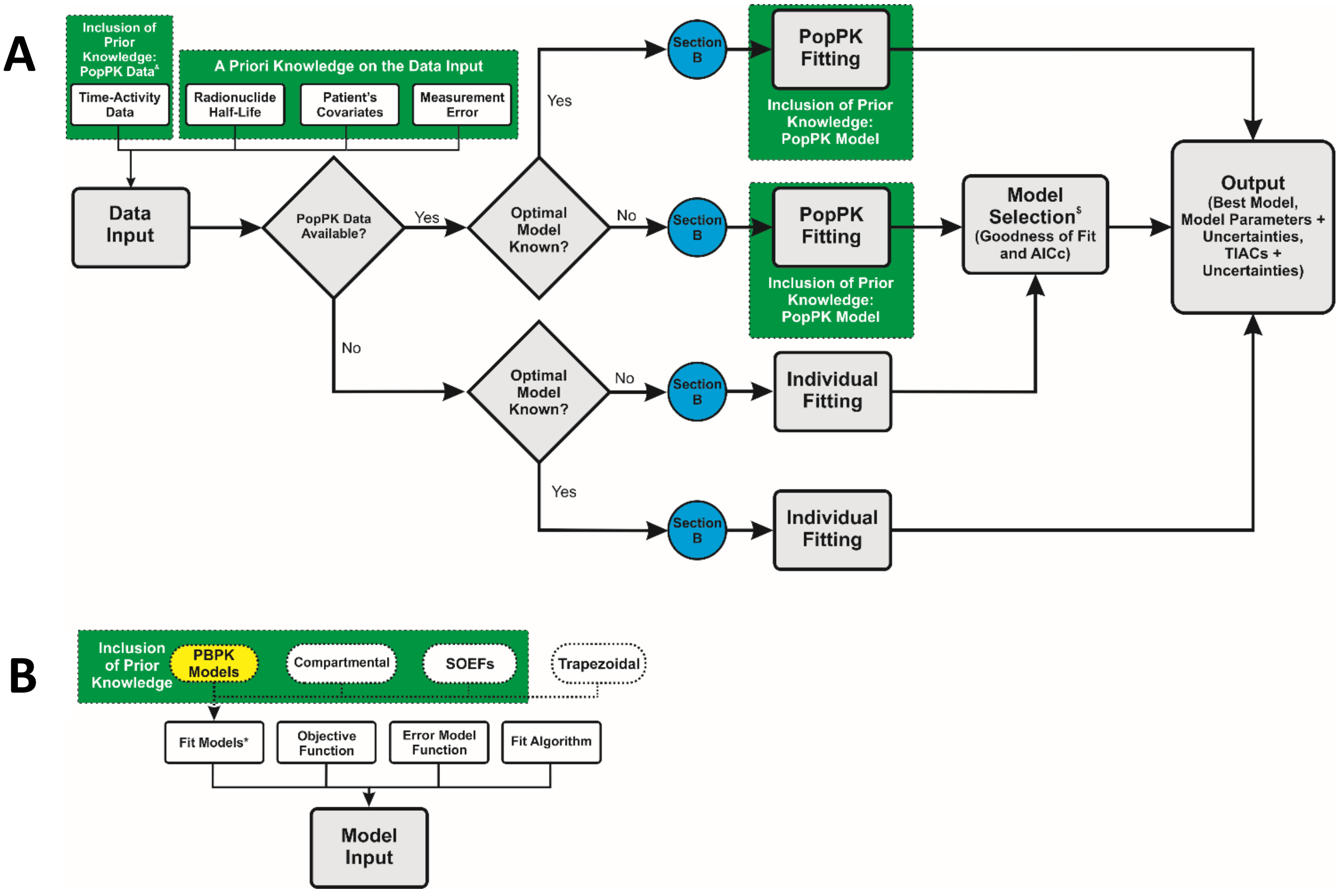

**Figure 1.** *(A)* Workflow for TAC fitting in RPT dosimetry. Input data include time-activity points, radionuclide half-life, patient covariates, and measurement errors. PopPK data can be incorporated when available. Individual fitting is used when population-level data are unavailable, while PopPK fitting is preferred when such data exist. Model selection is recommended when the optimal model structure is uncertain and should be guided by prior knowledge or the dataset itself. Reporting uncertainties in both model parameters and TIAC is important. *(B)* Key function inputs for TAC fitting include the fit function, objective function, error model, and fitting algorithm. Fit functions may range from simple (e.g., trapezoidal) to complex models (e.g., SOEFs, compartmental, PBPK models), depending on the level of prior knowledge. Model selection requires defining a candidate set of functions informed by physiological or empirical knowledge. For compartmental and PBPK models, where physiological knowledge is included, it is essential to evaluate the plausibility of fitted model parameter values. Parameter estimates should be compared with the range of directly measured physiological or pathological values reported in the literature to ensure the biological validity and reliability of the model.

In addition to improving the model structure, prior knowledge can also enhance the data processing step through the use of Population Pharmacokinetic (PopPK) modeling.[15,16,27-31] Unlike commonly used approaches that fit each patient's data independently, PopPK models leverage biokinetic information across a population, statistically characterizing both inter-individual and intra-individual variability in pharmacokinetic parameters. PopPK modeling enables more accurate function and parameter estimation than the individual fitting method, even when individual patient data are sparse, as is a frequent challenge in RPTs.[32,33] Compared to individual fitting, PopPK methods incorporate prior population-level knowledge into the estimation process, effectively increasing the information available for parameter estimation. This improves the data-to-parameter ratio, allowing for more robust model fitting. According to the fitting principles outlined in the EANM guidelines, the total number of data points (N) should exceed the number of estimated parameters (K).[13,34] When applying individual fitting, this requirement limits the number of parameters that can be reliably estimated (e.g., for N = 5 imaging time points, K should not exceed 4). In contrast, PopPK approaches, such as nonlinear mixed-effects modeling (NLMEM), combine data from all patients, including those with sparse data, into a unified modeling framework, thereby increasing the sample size N and permitting the fitting of more complex models with a higher number of identifiable parameters.[15]

Incorporating prior information, whether through PBPK-based function structure or PopPK-based data processing algorithms, strengthens the robustness of individualized TIAC estimation. These complementary strategies enhance the reliability of dosimetry, supporting personalized treatment planning and ultimately leading to better clinical outcomes. Fig. 1 presents the recommended steps for TAC analysis in RPT regarding the use of a priori knowledge. In the following, we summarize the current applications of PBPK and PopPK modeling in RPT and discuss future directions to optimize patient-specific dosimetry.

## Clinical Foundations of PBPK Modeling

PBPK modeling has gained some popularity in RPTs, where it aims to understand and optimize therapies, as well as improve patient safety.[23,26] In scenarios with limited imaging data, PBPK models provide reliable parameter estimates using fewer measurements. By incorporating in vitro and in vivo data, PBPK models facilitate the accurate prediction of tissue-specific drug distribution, assessment of the impact of tumor burden and dosing strategies on organ-specific radiation doses, and cross-species extrapolations, accommodating different administration routes and dosage levels.[35-43] Clinically, these models are also beneficial for subpopulations such as elderly or obese patients, and effectively quantify pharmacokinetic uncertainty.[44,45] In RPT, PBPK modeling is especially advantageous, as it overcomes logistical constraints inherent in limited imaging and sampling capabilities by simulating personalized biokinetics across various scanning times, patient profiles, and physiological scenarios.[10,38,46,47] Fig. 2 shows an example of the structure of a whole-body PBPK model.

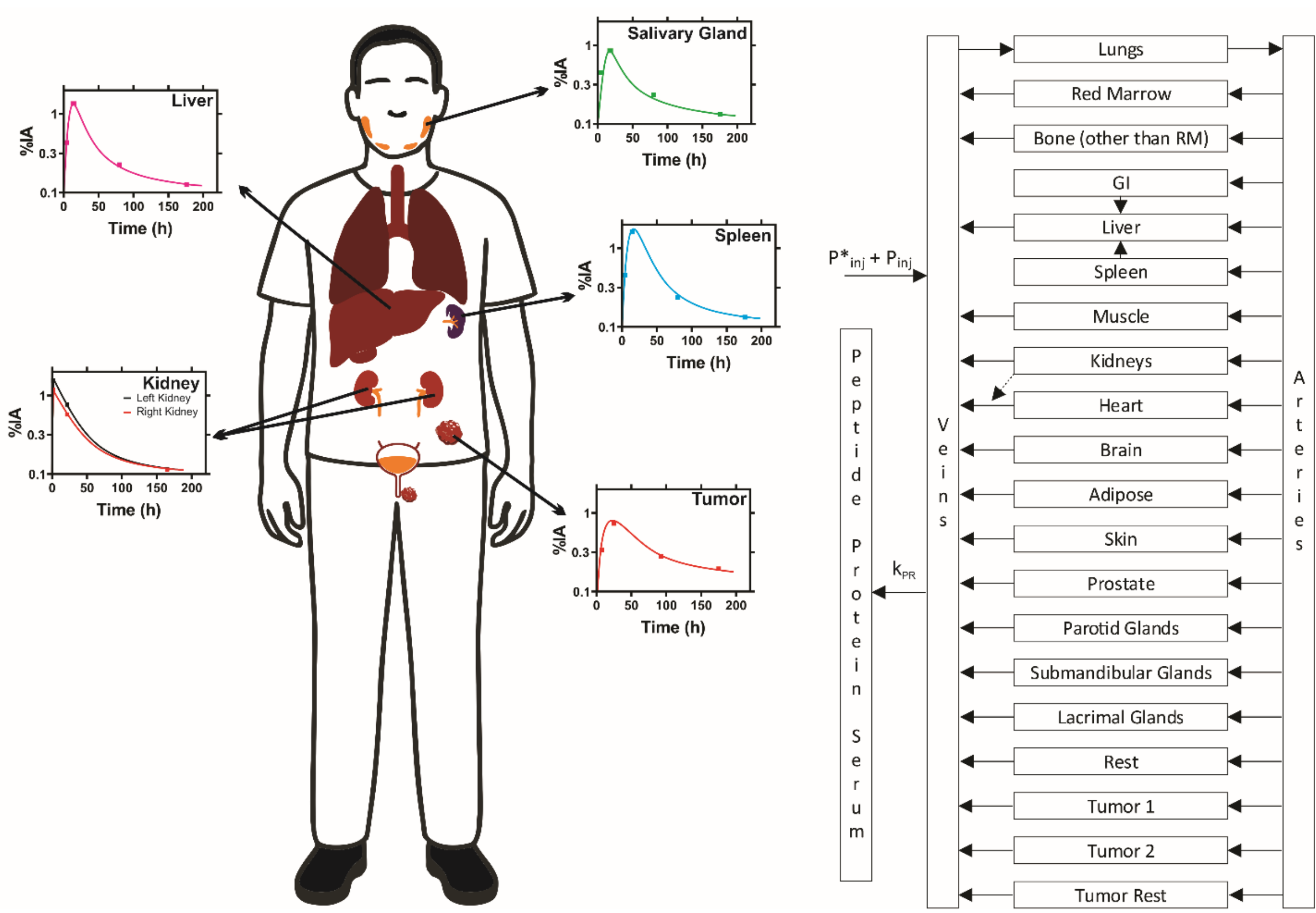

**Figure 2.** Whole-body physiologically based pharmacokinetic (PBPK) model structure for [$^{177}$Lu]Lu-PSMA. Each organ is represented as a rectangular compartment and interconnected via systemic blood circulation. Except for arteries, veins, the brain, and the peptide-protein serum compartment, all organs are subdivided into functional subcompartments. Renal clearance is modeled as the primary pathway of elimination. The "Peptide-Protein Serum" compartment accounts for serum protein-bound peptides. Due to the relatively low fraction of bound peptides, and to maintain model parsimony, this compartment is connected exclusively to the venous blood compartment. The models for the radiolabeled and unlabeled compounds are linked via competitive binding to a common molecular target.[48]

Despite these strengths, PBPK models present limitations that affect their reliability and applicability. Model accuracy critically depends on the quality and availability of input parameters, including physiological and drug-specific parameters such as body weight and dissociation constant, respectively; inaccuracies in these inputs may compromise model predictions.[49,50]

Additionally, constructing robust PBPK models requires extensive time, resources, and organ-specific pharmacokinetic data, which can be challenging to obtain.[50,51] For drugs with limited clinical data, achieving comprehensive model validation poses further difficulties. Although these models possess predictive capabilities beyond existing datasets, rigorous verification and validation through an iterative "verify-and-learn" approach are imperative.[52] Consequently, PBPK predictions should not be exclusively relied upon without thorough validation to maintain clinical robustness and reliability.

**Clinical Foundations of Population Pharmacokinetic Modeling**

PopPK is formally defined within a hierarchical modeling framework. At the first level, pharmacokinetic observations (e.g., drug concentrations) are described by individual-specific models, with parameters influenced by covariates and subject to intra-individual variability. At the second level, these individual parameters are treated as random variables whose distribution across the population (inter-individual variability) is modeled as a function of covariates. Two main approaches are used to estimate fixed effects and variability components: (1) the two-stage approach, suited for data-rich settings, and (2) the NLMEM approach, which is particularly advantageous in data-sparse scenarios. The latter is the primary focus of modern PopPK analyses due to its ability to leverage sparse and unbalanced datasets.

The traditional two-stage approach to PK analysis begins by estimating individual PK parameters using nonlinear regression on dense data.[53-57] In the second stage, these individual estimates are summarized to obtain population-level statistics, such as means, variances, and covariances. Covariate analysis may also be conducted using classical statistical methods (e.g., regression or ANOVA). While this method can provide unbiased mean estimates, it tends to overestimate variability, especially under realistic conditions. Refinements such as the global two-stage approach have been introduced to correct for bias and account for differences in data quality and quantity. Despite its historical role in drug development, the two-stage method has largely been supplanted by more advanced techniques like nonlinear mixed-effects modeling in modern PopPK analyses.[58]

In sparse-data scenarios where individual parameter estimation is infeasible, NLMEM provides a robust single-stage solution. This approach evolved from the need to study pharmacokinetics and pharmacodynamics in patients. Unlike the two-stage method, NLMEM treats the population as the unit of analysis and can accommodate sparse, unbalanced, and observational data. It estimates both fixed effects (population means) and random effects (inter-individual variability), providing a comprehensive description of parameter distributions and covariate relationships.[59,60] By directly fitting all individual concentration data, NLME preserves individual variability while enabling population-level inference, even when sampling is limited. As such, it is now the preferred method for PopPK analysis in drug development and clinical applications. In RPT, PopPK modeling with NLMEM has proven advantageous due to its ability to handle sparse datasets, a frequent limitation in RPT studies (Fig. 3).[17,33,61] Commonly used individual fitting methods typically require more measurements than can be safely achieved in RPT. By leveraging population data pooled from many patients, PopPK modeling increases the ratio between the total number of data points and the number of model parameters through simultaneous fitting, thereby enhancing the robustness of parameter estimates. This feature also supports the objective model selection between pharmacokinetic models. Recent studies have demonstrated that PopPK modeling, coupled with population-based model selection (PBMS), outperforms individual-based model selection and also several commonly used SOEFs in the context of RPT.[29,33] This methodology offers a statistically supported framework for deriving reference pharmacokinetic models, which are currently lacking in RPT dosimetry. PopPK analysis offers substantial clinical advantages, particularly its flexibility in supporting diverse study designs, including routine clinical settings.[62] PopPK modeling provides a powerful framework for systematically identifying and quantifying covariates that contribute to variability in radiopharmaceutical distribution, effectively capturing both interindividual and intraindividual differences in pharmacokinetics.[63]

Despite these strengths, PopPK approaches possess several limitations. Reliable PopPK models typically require a substantial patient population, which can pose challenges in patient recruitment and data collection.[64] Additionally, PopPK modeling involves the application of advanced pharmacostatistical techniques and the use of high-quality, rigorously validated datasets, making the process both time-intensive and resource-demanding. Model development processes are complex and time-consuming, which can hinder interpretation and accessibility, particularly among clinicians or researchers without specialized training in pharmacometrics. Furthermore, PopPK analyses can struggle with missing covariate data, particularly when covariates are incompletely recorded, thereby affecting model robustness and applicability in covariate model development.

In the next section, we highlight examples such as PBPK models for antibody preloading, PET-informed TIAC prediction, and treatment planning beyond fixed-activity protocols, along with PopPK models that identify covariates, optimize sampling schedules, and characterize inter-individual variability. More recently, integrated PBPK–PopPK approaches and simplified PK models have also shown promise for accurate absorbed dose estimation and personalized treatment planning.

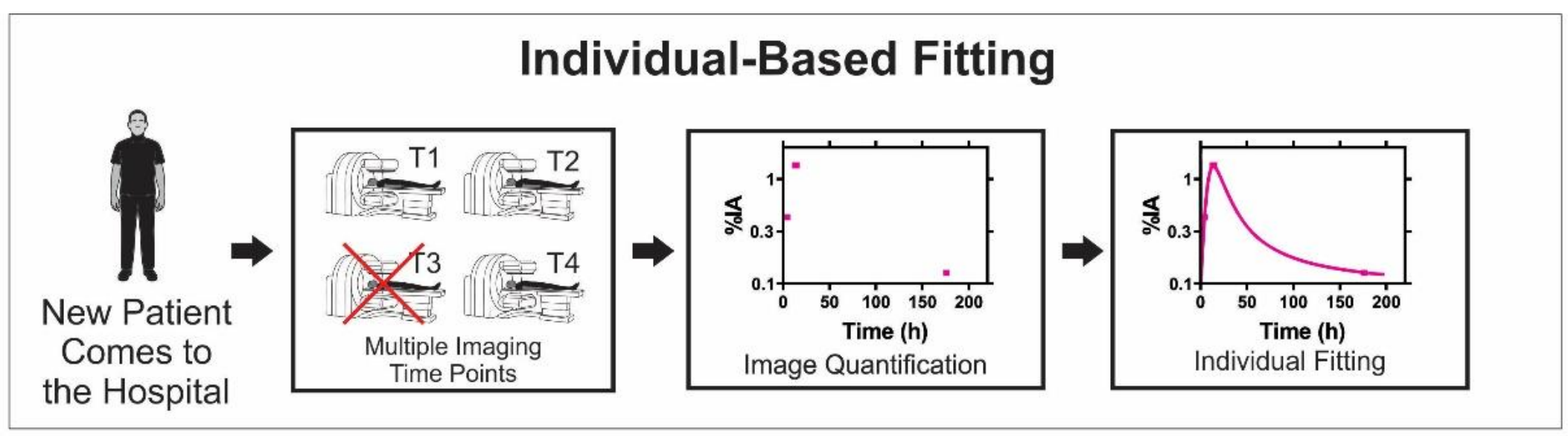

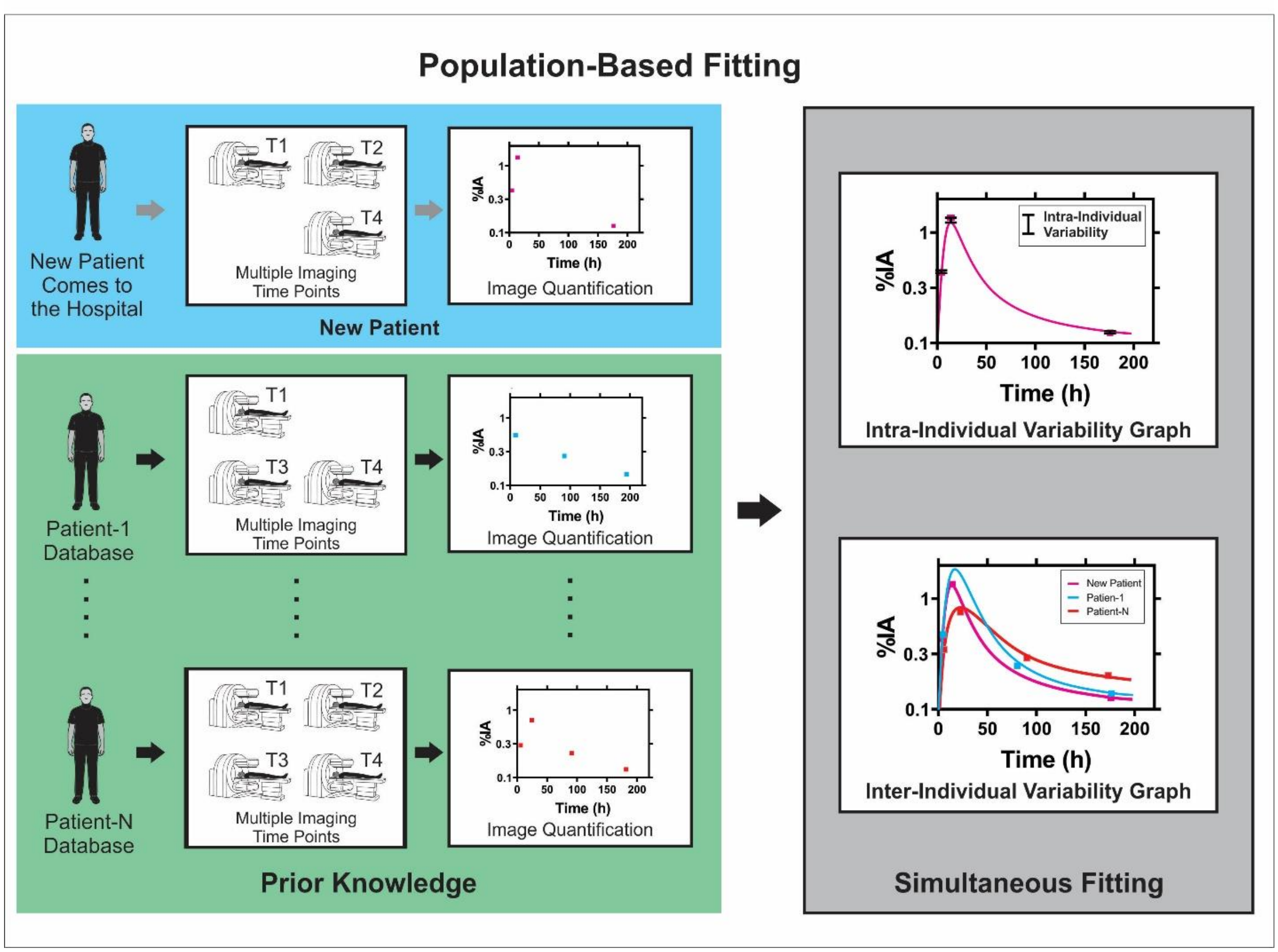

**Figure 3.** Schematic representation of population pharmacokinetic (PopPK) modeling using nonlinear mixed-effects modeling (NLMEM). Instead of fitting each patient's data separately, all patient data are modeled simultaneously. A new patient with sparse data (blue) can be incorporated alongside existing clinical data (green), enabling robust parameter estimation. The model outputs include population fixed effects, inter- and intra-individual variability, individual parameter estimates, and associated uncertainties (grey).[15]

**Examples of the implementation of PBPK and PopPK modeling in RPT**

Virtual theranostics trials using PBPK modeling hold tremendous potential for optimizing RPTs, including the development of digital twins to simulate imaging schedules, dosing strategies, and physiological variability.[40,65-70,] The first implementation of a PBPK model in RPT demonstrated its potential for individualizing the optimal antibody preload in radioimmunotherapy using the YAML568 anti-CD45 antibody.[23] Another study by Hardiansyah et al. demonstrated that integrating PET data with PBPK modeling enables the accurate prediction of TIACs for therapeutic agents, such as [$^{90}$Y]Y-DOTA-TATE.[65,66] Likewise, Jiménez-Franco et al. demonstrated that PBPK-informed treatment planning can surpass fixed-activity protocols by enhancing tumor control probability while maintaining organ dose limits. Further, they identified key physiological thresholds required for effective therapy.[67,69] There is a growing number of studies focused on simulating and analyzing radiopharmaceutical therapies to optimize treatment strategies.[71-73] Collectively, these studies underscore the potential of PBPK-based *in silico* modeling to improve understanding of RPTs (e.g., how injection schedules or properties of different radiopharmaceuticals (e.g., albumin-binding) can impact absorbed doses) and to help enable predictive dosimetry, guide personalized treatment planning to improve therapeutic outcomes and facilitates the advancement of novel radiopharmaceuticals.

Pharmacokinetic (PK) and pharmacodynamic (PD) modeling are essential tools for optimizing dosing strategies and enhancing therapeutic outcomes in RPTs. The former models what "the body does to the drug", which PBPK and PopPK models aim for, while the latter models what "the drug does to the body", which in the case of RPTs is represented by radiobiological models (see the contribution to this special issue by Yusufaly et al. on computational radiobiology.[74] Several studies have demonstrated the value of using PopPK and PBPK models in characterizing the behavior of $^{177}$Lu-labeled agents and explaining variability in treatment response. Puszkiel et al. used a PopPK model to assess the impact of amino acid co-infusion on [$^{177}$Lu]Lu-DOTA-TATE pharmacokinetics, revealing increased elimination (k10) and high variability, which correlated with lymphopenia.[22] Kletting et al. developed a PBPK/PD model to predict PSMA-positive tumor volume post-therapy, achieving close agreement with measured outcomes and identifying radiosensitivity parameters.[75] Siebinga et al. employed a five-compartment PopPK/PD model for [$^{177}$Lu]Lu-PSMA-I&T, linking SPECT-based tumor uptake with PSA dynamics and uncovering key covariates like renal function and tumor burden that affect treatment response.[76]

Recent studies have demonstrated that PBMS using NLMEM PopPK modeling can serve as a robust reference framework in RPT, offering improved accuracy, consistency, and reliability for model-based dosimetry and treatment planning.[17,29,33] By leveraging population-level data, PBMS NLMEM ensures consistent model structure selection and has demonstrated superior performance over individual-based model selection, particularly in single-time-point dosimetry applications.[16,17,29] The PopPK approach has also proven to be a powerful methodology for optimizing sampling schedules to enable accurate absorbed dose estimation in RPT dosimetry.[17,29,31]

PopPK modeling has also been shown to be a powerful approach for quantifying inter-individual variability in radiopharmaceutical kinetics by identifying relevant covariates. Van Rij et al. developed a two-compartment PopPK model for [$^{11}$C]flumazenil PET data from 51 patients, revealing that clearance was 20% lower in epilepsy patients compared to those with depression, and the central volume of distribution increased with body weight (0.55% per kg).[77] These covariate effects enabled the design of an optimized sampling schedule using only two time points (30 and 60 min), accurately estimating AUC with minimal bias. Similarly, Siebinga et al. applied a five-compartment PopPK model to [$^{177}$Lu]Lu-PSMA I&T SPECT data in 76 patients with mCRPC, integrating covariates such as renal function and tumor volume, and linking tumor uptake to PSA dynamics in a PK/PD framework.[78] Incorporating covariate data enhances the predictive accuracy of PopPK models, facilitates individualized dosing strategies, and supports the optimization of sampling protocols in radiopharmaceutical therapy RPT.

Some studies have demonstrated the effectiveness of integrating PopPK with PBPK modeling in therapies such as radioimmunotherapy (RIT) and peptide receptor radionuclide therapy (PRRT).[25,79,80] By applying PopPK within PBPK models, predictions of therapeutic biodistributions become more accurate and less resource-intensive, improving the accuracy of single-time-point (STP) dosimetry calculations without requiring multiple time-point imaging. For instance, population-based parameters significantly improved the prediction accuracy of therapeutic TIACs in [$^{90}$Y]Y-labeled anti-CD66 antibody RIT.[81] Another study demonstrated that the implementation of PBPK and PopPK modeling could enhance STP dosimetry by enabling accurate and precise estimation of absorbed doses in various organs.[25,80] Simpler pharmacokinetic (PK) models that incorporate prior knowledge can also be employed for absorbed dose calculations. For example, a recent study by Zaid et al. demonstrated the use of a simplified model to optimize the number of treatment cycles in RPTs.[82] While such models may lack the structural complexity needed to capture detailed physiological processes, they can be effective for predictive dosimetry in clinical decision-making. Therefore, it is essential to include comparative evaluations between these simplified approaches and more comprehensive modeling techniques to assess their relative strengths, limitations, and suitability for different clinical scenarios. Notably, these simplified models can be further enhanced by integrating PopPK modeling, which enables improved parameter estimation even in data-sparse settings.

Together, these advances show how PBPK and PopPK modeling can move beyond population-level characterization to support individualized treatment planning in RPT. By guiding patient-specific dosing strategies, these approaches have the potential to enhance therapeutic efficacy while reducing toxicity.

**Future directions of PopPK and PBPK models in RPT**

The future of PK modeling in RPTs, encompassing both PopPK and PBPK approaches, will be significantly shaped by technological innovation and the diversification of radiopharmaceutical agents. As highlighted in recent developments, virtual theranostic trials (VTTs) offer a promising framework for evaluating and optimizing RPT strategies *in silico*. Platforms such as the one proposed by Fedrigo et al. provide a foundation for implementing VTT workflows across diverse patient scenarios.[83] The predictive power and clinical relevance of VTTs can be further enhanced by integrating PBPK and PopPK modeling, enabling mechanistic simulation and individualized parameter estimation, even in data-sparse settings.

The integration of artificial intelligence (AI) presents a promising opportunity to complement and enhance the performance of PBPK and PopPK models. AI techniques can support improved covariate selection, parameter estimation, and absorbed dose optimization. By combining these computational approaches, it becomes possible to develop highly adaptive models that better capture complex inter-individual variability, ultimately enabling more accurate and personalized treatment strategies across heterogeneous patient populations.

Another future direction involves the development of multiscale PK models that bridge molecular, cellular, tissue, and whole-body pharmacokinetics. As new molecular targets, radionuclide carriers, and therapeutic isotopes are introduced into radiotheranostic applications, PopPK and PBPK models must evolve to capture the biological complexities associated with these innovations. For instance, unconventional tracers such as $^{18}$F-proline, a marker of collagen synthesis rather than receptor–ligand binding,[84] highlight the need for model frameworks flexible enough to represent atypical biological pathways. Multiscale approaches will allow the integration of tumor microenvironment heterogeneity, receptor dynamics, and physiological adaptations over time, providing a deeper mechanistic understanding that can enhance the precision of absorbed dose predictions and therapeutic efficacy.

The implementation of PopPK and PBPK models in RPT faces considerable challenges, particularly the high level of expertise required for model development, validation, and clinical application. The complexity of these models, coupled with the limited availability of trained personnel, remains a key barrier to their broader use. Additional obstacles include extensive data requirements, parameter uncertainty, software limitations, and difficulties with clinical integration. Addressing these challenges will require expanding and standardizing educational opportunities, such as accessible hands-on workshops, while simultaneously advancing tools and infrastructure that lower the technical threshold for end-users. For widespread clinical adoption, streamlined PK modeling tools and supportive regulatory frameworks will be critical.[58,85] Future efforts are expected to focus on the development of user-friendly software platforms, standardized modeling protocols, and automated workflows that integrate seamlessly with imaging systems and clinical data management. These platforms should conceal methodological complexity, allowing clinicians to benefit from advanced modeling without needing detailed technical expertise.

In parallel, establishing clear regulatory guidelines for model validation and application in radiopharmaceutical development will be essential to ensure model transparency, reproducibility, and acceptance by healthcare authorities. These advancements will position PopPK and PBPK modeling as important pillars in the future of personalized RPTs.

## Conclusion

PBPK models offer a mechanistic framework for simulating radiopharmaceutical kinetics based on physiological parameters, while supporting extrapolations across populations, dosing regimens, and clinical scenarios. Complementary to this, PopPK models leverage clinical data to characterize pharmacokinetic variability and identify covariates critical for individualized dose optimization, even under sparse sampling conditions. Future developments are expected to focus on creating hybrid PBPK-PopPK models and incorporating pharmacodynamic endpoints invoking radiation biology models to better link absorbed doses with therapeutic outcomes. The adoption of artificial intelligence, improved imaging technologies, and digital health tools will further enhance model performance and facilitate clinical implementation. Ultimately, these modeling approaches are expected to translate into direct patient benefits by enabling reduced absorbed doses to healthy organs and improved tumor control through model-based personalization of therapy. Together, PBPK and PopPK approaches can help establish the next generation of precision dosimetry and patient-tailored RPTs.

## Acknowledgements

We appreciate helpful discussions with Dr. Hamid Abdollahi about the concept of virtual theranostic trials (VTTs). DH was supported by Hibah PUTI Q1 2025 from Universitas Indonesia with grant number PKS-200/UN2.RST/HKP.05.00/2025. GG was supported by the DAAD with the resources of the German Federal Ministry for Economic Cooperation and Development (BMZ) with project number 57657089.